\documentclass[prd,floatfix,nofootinbib,preprintnumbers]{revtex4}
\usepackage{graphicx}
\usepackage{subfigure}

\begin{document}
\title{ Neutral Higgs boson contributions to CP asymmetry of
$B \to \Phi K_S$ in MSSM }

\author{Jian-Feng Cheng$^a$,  Chao-Shang Huang$^a$, and Xiao-Hong Wu$^{b,c}$}
\affiliation{
 $^a$ Institute of Theoretical Physics, Academia Sinica, P. O. Box 2735,
             Beijing 100080,  China\\
 $^b$ Department of Physics, Peking University, Beijing 100871,
             China \\
 $^c$ CCAST (World Lab.), P.O. Box 8730, Beijing 100080, China
}
%\maketitle

\renewcommand{\textfraction}{0}
\def\llgm{\left\lgroup\matrix}
\def\rrgm{\right\rgroup}

\def\beq{\begin{equation}}
\def\eeq{\end{equation}}
\def\bea{\begin{eqnarray}}
\def\eea{\end{eqnarray}}
\def\nnb{\nonumber}
\def\rt{\right}
\def\dgr{\dagger}
\def\lt{\left}
\def\Tr{\textrm Tr}
\def\ncp{\nu_e + d \to n+p+\nu_e}
\def\tY{\tilde Y}
\def\tal{\tilde \alpha}
\newcommand{\wti}{\widetilde}
\newcommand{\gsim}{\lower.7ex\hbox{$\;\stackrel{\textstyle>}{\sim}\;$}}
\newcommand{\lsim}{\lower.7ex\hbox{$\;\stackrel{\textstyle<}{\sim}\;$}}
\renewcommand{\thefootnote}{\fnsymbol{footnote}}

\newcommand\epjc[3]{Eur.\ Phys.\ J.\ C {\bf #1}, #3 (#2)}
\newcommand\ijmpa[3]{Int.\ J.\ Mod.\ Phys.\ A {\bf #1} (#2) #3}
\newcommand\jhep[3]{J.\ High Ener.\ Phys.\ {\bf #1} (#2) #3}
\newcommand\npb[3]{Nucl.\ Phys.\ B {\bf #1} (#2) #3}
\newcommand\npps[3]{Nucl.\ Phys.\ B (Proc.\ Suppl.) {\bf #1} (#2) #3}
\newcommand\plb[3]{Phys.\ Lett.\ B {\bf #1} (#2) #3}
\newcommand\prep[3]{Phys.\ Rep.\ {\bf #1} (#2) #3}
\newcommand\sjnp[3]{{Sov.\ J.\ Nucl.\ Phys.\ }{\bf#1} (#2) #3}
\newcommand\yf[3]{{Yad.\ Fiz.\ }{\bf#1} (#2) #3}
\newcommand\zpc[3]{Z.\ Phys.\ C {\bf #1} (#2) #3}
\newcommand{\hepph}[1]{{\tt hep-ph/#1}}
\newcommand{\heplat}[1]{{\tt hep-lat/#1}}
\newcommand{\hepex}[1]{{\tt hep-ex/#1}}

\begin{abstract}
 We have studied the neutral Higgs boson (NHB) contributions to the pure
penguin process $B\to \phi K_S$ in MSSM with middle and large
$\tan\beta$ (say, $>$ 8). We show that the $\alpha_s$ order
hadronic matrix elements of NHB induced operators can make sizable
 effects on both the branch ratio and time dependent
CP asymmetry $S_{\phi K}$. Under the all relevant experimental
constraints, the Higgs mediated contributions to $S_{\phi K}$
alone can provide a significant deviation from SM and, in
particular, lead to a negative $S_{\phi K}$ which is reported by
BaBar and Belle.

\end{abstract}
\maketitle
%\newpage

Among charmless hadronic B decays, the rare decay $B \to \Phi K_S$ is one of channels
which provide a powerful testing ground for new physics. The reason is simply because
it has no tree level contributions in the standard model (SM). The decay has been studied
in SM and the predicted branching ratio (Br) in the perturbative QCD framework is
 $(3.5-4.3)\times 10^{-6}$~\cite{hmw} by using the BBNS approach~\cite{bbns1,bbns2} or
about $10\times 10^{-6}$~\cite{mish} by using the H.-n. Li et al's
approach~\cite{lhn}, which is somewhat smaller than or consistent
with the measured value $(8.4^{+2.5}_{-2.1})\times 10^{-6}$, the
current world average from BaBar~\cite{babarphi} and
Belle~\cite{bellephi}.

The recently reported measurements of time dependent CP
asymmetries in $B \to \Phi K_S$ decays by BaBar \cite{babarphi}
%\begin{eqnarray}
%\label{eq:babar} \sin (2 \beta (\Phi K_S))_{BaBar}=-0.19
%^{+0.52}_{-0.50} \pm 0.09
%\end{eqnarray}
and Belle \cite{bellephi} %$\begin{eqnarray} \label{eq:belle} \sin
%(2 \beta (\Phi K_S))_{Belle}=-0.73 \pm 0.64 \pm 0.18
%\end{eqnarray}
lead to the error weighted average
\begin{equation}
\label{eq:ave}S_{\phi K}= \sin (2 \beta (\Phi K_S))_{ave}=-0.39
\pm 0.41 \;
\end{equation}
with errors added in quadrature\footnote{2003 data is $-0.96\pm
0.50^{+0.09}_{-0.11}$ by Belle~\cite{belle03} and it is $+0.45 \pm
0.43\pm 0.07$ by BaBar~\cite{babar03}.}.  %(for the definition of
%$S_{\phi K}$, see, e.g., Ref.~\cite{kane}).

Here the time-dependent $CP$-asymmetry $S_{\phi K}$ is defined by
\begin{equation}
a_{\phi K}(t) = - C_{\phi K} \cos(\Delta M_{B_d^{0}} t) + S_{\phi
K} \sin(\Delta M_{B_d^{0}}t),
\end{equation}
where
\begin{equation}
C_{\phi K} = \frac{1-|\lambda_{\phi k}|^2}{1+|\lambda_{\phi k}|^2}
\,, \ \ \ \ \ \ \ \ \ \ S_{\phi K} = \frac{2\,\mathrm{Im}
\lambda_{\phi k}}{1+|\lambda_{\phi k}|^2} \; ,
\end{equation}
with $\lambda_{\phi k}$ being
\begin{equation}
\lambda_{\phi k} =
\left(\frac{q}{p}\right)_B\frac{\mathcal{A}(\bar{B}^0_d\rightarrow
\phi K_S)} {\mathcal{A}(B^0_d\rightarrow \phi K_S)} \; .
\end{equation}

In the SM the above asymmetry is related to that in $B \to J/\Psi
K_S$ \cite{Grossman:1996ke} by
\begin{equation}
\label{eq:diff}  \sin (2 \beta (\Phi K))= \sin (2 \beta (J/\Psi
K)\!) \! + \! O(\lambda^2 \!)
\end{equation}
where $\lambda \simeq 0.2$ appears in Wolfenstein's
parameterization of the CKM matrix and
\beq \sin (2 \beta (J/\Psi
K_{S,L}))_{world-ave}=0.734 \pm 0.054.
\eeq
Therefore,
(\ref{eq:ave}) violates the SM at the 2.7 $\sigma$ deviation.

Obviously, the impact of these experimental results on the
validity of CKM and SM is currently statistics limited. However,
they have attracted much interest in searching for new
physics~\cite{dat,kk,kane}.

In the letter we inquire the possibility to explain the 2.7
$\sigma$ deviation in R-parity conservative supersymmetric models,
e.g., the minimal supersymmetric standard model (MSSM). There are
mainly two new contributions arising from the QCD and
chromomagnetic penguins and neutral Higgs boson (NHB) penguins
with the gluino and squark propagated in the loop.
 The QCD and chromomagnetic penguin contributions to
$b\rightarrow ss\bar s$ have been analyzed in Refs.\cite{kk,kane}.
The NHB penguin contributions have been discussed in
Ref.\cite{kane}. However, the conclusion on NHB contributions,
$S_{\phi K}$ can not smaller than 0.71, in Ref.\cite{kane} is
valid only in some special cases, e.g., the extended minimal
flavor violation (MFV) scenarios\cite{kane} with the naive
factorization of hadronic matrix elements of operators and the
decoupling limit in MSSM, and it is not valid in (general) MSSM,
as we shall show. We concentrate on the Higgs penguin
contributions in MSSM in the letter. An interesting fact is that
the Higgs penguins contribute to $B\rightarrow \phi K_S$ but not
$B\to J/\psi K_S$ at the $\alpha_s$ order of hadronic matrix
element calculations due to the non-match of color and flavor. At
the tree level of hadronic matrix element calculations (i.e., the
naive factorization) their contributions to the branch ratio of
$B\to J/\psi K_S$ (the time dependent CP asymmetry $S_{J/\psi K}$)
are very small (negligible). Therefore, we do not need to worry
about the effects of the Higgs penguins on $S_{J/\psi K}$. It is
shown that the Higgs mediated contributions in the case of middle
and large $\tan\beta$ (say, $\geq 8$) are important under the
constraints from the experimental bounds of $B_s\rightarrow
\mu^+\mu^-$ and $\Delta M_s$ as well as $b\to s \,\gamma, s\,g$,
$B\to X_s \mu^+\mu^-$ and consequently can have significant
effects on $B \to \Phi K_S$ due to the $\tan^2\beta$ enhancement
of Wilson coefficients of the NHB induced operators in
MSSM~\cite{beku,hw}.
 %We discuss
%the impact of the uncertainty of hadronic matrix elements on Br
%and CP asymmetry.
Our results show that the Higgs mediated contributions to $S_{\phi
K}$ alone can provide a significant deviation from the SM and %for
%$\mu\sim m_b/2$ ($\mu$ is the characteristic scale of the process)
 a possible explanation of the 2.7 $\sigma$ deviation, satisfying
all the relevant experimental constraints.

The effective Hamiltonian induced by neutral Higgs bosons can be
written as
\begin{equation}\label{Heffn}
   {\cal H}_{\rm eff}^{neu} = \frac{G_F}{\sqrt2}(-\lambda_t)
    \!\sum_{i=11,\dots, 16}\! (C_i\,Q_i + C_i^{\prime}\,Q_i^{\prime})
   + \mbox{h.c.} \,,
\end{equation}
where $Q_{11}$ to $Q_{16}$, the neutral Higgs penguins operators,
are given by
\begin{eqnarray}
Q_{11(13)} &=& (\bar s\, b)_{S+P} \sum_q\,{m_q\over m_b} (\bar q\,
q)_{S-(+)P} \,, \nonumber\\
Q_{12(14)} &=& (\bar s_i \,b_j)_{S+P}
 \sum_q\,{m_q\over m_b}(\bar q_j \,q_i)_{S-(+)P} \,, \nonumber\\
%Q_{13} &=& (\bar s \,b)_{S+P} \sum_q\,{m_q\over m_b}
%  (\bar q\, q)_{S+P} \,, \nonumber\\
%Q_{14} &=& (\bar s_i \,b_j)_{S+P} \sum_q\,{m_q\over m_b}
%   (\bar q_j\, q_i)_{S+P} \,, \nonumber\\
Q_{15} &=& \bar s \,\sigma^{\mu\nu}(1+\gamma_5) \,b
\sum_q\,{m_q\over m_b}
    \bar q\, \sigma_{\mu\nu}(1+\gamma_5)\,q \,, \nonumber\\
Q_{16} &=& \bar s_i \,\sigma^{\mu\nu}(1+\gamma_5) \,b_j \sum_q\,
    {m_q\over m_b} \bar q_j\, \sigma_{\mu\nu}(1+\gamma_5) \,q_i
    \, ,
\end{eqnarray}
where $(\bar q_1 q_2)_{S\pm P}=\bar q_1(1\pm\gamma_5)q_2$
\footnote{Strictly speaking, the sum over q in expressions of
$Q_i$ should be separated into two parts: one is for q=u, c, i.e.,
upper type quarks, the other for q=d, s, b, i.e., down type
quarks, because the couplings of upper type quarks to NHBs are
different from those of down type quarks. In the case of large
$\tan\beta$ the former is suppressed by $\tan^{-1}\beta$ with
respect to the latter and consequently can be neglected. Hereafter
we use, e.g., $C_{11}^c$ to denote the Wilson coefficient of the
operator $Q_{11}= (\bar s\, b)_{S+P} \,{m_c\over m_b} (\bar c\,
c)_{S-P}$.}.
 The operators $Q_i^\prime$s are obtained from the unprimed operators
$Q_i$s by exchanging $L\leftrightarrow R$.

There are new sources of flavor violation in MSSM. Besides the CKM
matrix, the $6\times 6$ squark mass matrices are generally not
diagonal in flavor (generation) indices in the super-CKM basis in
which superfields are rotated in such a way that the mass matrices
of the quark field components of the superfields are diagonal.
This rotation non-alignment in the quark and squark sectors can
induce large flavor off-diagonal couplings such as the coupling of
gluino to the quark and squark which belong to different
generations. These couplings can be complex and consequently can
induce CP violation in flavor changing neutral currents (FCNC). It
is well-known that the effects of the primed counterparts of usual
operators are suppressed by ${m_s}/{m_b}$ and consequently
negligible in SM because they have the opposite chiralities.
However, in MSSM their effects can be significant, since the
flavor non-diagonal squark mass matrix elements are free
parameters. We have calculated Wilson coefficients of $Q_i$ and
$Q_i^\prime$ coming from the gluino-squark loop contributions at
the $m_w$ scale  and their explicit expressions are given in
Ref.\cite{hw}. We evolve the Wilson coefficients to the low scale
$\mu\sim m_b$ using the one loop QCD anomalous
dimensions~\cite{adm,bghw,hk}.

The effective Hamiltonian (\ref{Heffn}) results in the following
matrix element for the decay $B\rightarrow \phi K_s$
\begin{equation}\label{Topn}
   \langle K_s \phi|{\cal H}_{\rm eff}^{neu}| B\rangle
   = \frac{G_F}{\sqrt2} \sum_{p=u,c} \lambda_p\,
   \langle K_s \phi|{\cal T}_p^{\rm neu}| B\rangle \,,
\end{equation}
where ${\cal T}_p^{neu}$ is given by
\begin{eqnarray}\label{tneu}
{\cal T}^{\rm neu }_p &&= a_4^{neu} ( {\bar s}\,  b)_{\rm V-A}
         \otimes ( {\bar s}\,  s )_{\rm V-A} \nonumber\\
&& + {m_s \over m_b} \big[ -{1\over 2}a_{12} ({\bar s}\,  b)_{\rm V+A}
\otimes ( {\bar s}\,  s )_{\rm V-A} \nonumber\\
&&-{1\over 2} a_ { 12}^\prime ( {\bar s}\, b)_{\rm V-A}
         \otimes ( {\bar s}\,  s )_{\rm V+A} \nonumber\\
&&+ {4 m_s\over m_b}\,(a_{16}+a_{16}^\prime) ( {\bar s}\, b)_{\rm V-A}
         \otimes ( {\bar s}\,  s )_{\rm V-A} \big]\, .
\end{eqnarray}
In Eq.(\ref{tneu})
\begin{eqnarray}\label{ai}
\vspace{0.1cm}
   a_4^{neu} &=& {C_F\alpha_s\over 4\pi}\,{P_{\phi,2}^{neu}\over N_c}
\,,\nonumber\\
   a_{12} &=& C_{12} + {C_{11}\over N_c} \left[ 1
    + {C_F \alpha _s \over 4 \pi  } \left( -V^\prime
     -{4 \pi  ^2 \over N_c} H_{\phi K} \right) \right]\,,\nonumber\\
   a_{16} &=&  C_{16} + \displaystyle{\frac{C_{15}}{N_c} }\,,
\end{eqnarray}
with $P_{\phi,2}^{neu} $ being
\begin{eqnarray}\label{P2phi}
   P_{\phi,2}^{neu} &=&  -{1\over 2}
 ( C_{11} + C_{11}^\prime ) \nonumber\\
   &&\ \ \times  \left[ {m_s \over m_b}  \left( {4\over 3} \ln {
m_b \over \mu } - G_\phi (0)\right)  +\left( { 4\over 3} \ln { m_b
\over \mu } - G_ \phi (1) \right) \right] \nonumber \\
  && + ( C_{13} + C_{13}^\prime ) \left[ -2 \ln { m_b \over \mu }
       \,G^0_\phi - GF_ \phi (1) \right] \nonumber\\
  &&- 4 (C_{ 15} +C_{15}^\prime) \left[ \left( -{1\over 2}
-2 \ln { m_b \over \mu }\right) G^0_\phi - GF_\phi (1) \right] \nonumber\\
 &&- 8  (C_{16}^c + C_{16}^{\prime c})
 \bigg[ \left( {m_c \over m_b } \right)  ^2 \,
  \left( - 2 \ln { m_b \over \mu }\,G^0_\phi - GF_ \phi(s_c )
  \right)\bigg]
\nonumber\\
&& - 8  (C_{16}+C_{16}^\prime) \left[
  -2 \ln { m_b\over \mu }\,G^0_\phi
 - GF_\phi(1)
 \right]
% \nonumber\\
% &&- 8  (C_{16}^c + C_{16}^{\prime c})
% \bigg[ \left( {m_c \over m_b } \right)  ^2 \,
%  \left( - 2 \ln { m_b \over \mu }\,G^0_\phi - GF_ \phi(s_c )
%  \right)\bigg]
%\nonumber\\
    % &&\mbox{}- 2 C_{8g}^{\rm eff} \int_0^1 \frac{dx}{1-x}\,
    %\Phi_{\phi}(x) \,,\\
\end{eqnarray}
where
\begin{eqnarray}
&&G^0_\phi = \int_0^1 \frac{dx}{\bar x}\,
    \Phi_{\phi}(x)\, ,~~~~~~~GF(s,x) = \int^1_0 dt \ln \big[ s-x \,t {\bar t} \big]\, ,
   \nonumber \\
%\end{equation}
%\begin{equation}
&& GF_\phi (s) = \int^1_0 dx { \Phi _\phi (x)\over \bar x}
\, GF(s-i\, \epsilon , \bar x)\,
\end{eqnarray}
with $\bar x =1-x$ and $\Phi_{\phi}(x)= 6 x \bar x$ in the
asymptotic limit of the leading-twist distribution amplitude. In
calculations we have set $m_{u,d}=0$ and neglected the terms which
are proportional to $m_s^2/m_b^2$ in Eq.(\ref{P2phi}). We have
included only the leading twist contributions in Eq.(\ref{ai}).
In Eq.(\ref{tneu}) $a_i^\prime$ is obtained from $a_i$ by
substituting the Wilson coefficients $C_j^\prime$s for $C_j$s.
Here we have used the BBNS approach~\cite{bbns1,bbns2} to
calculate the hadronic matrix elements of operators. Notations
 not explicitly defined are the same as those in
Ref.\cite{bbns2}.

Before numerical calculations a remark is in place. From Eq.
(\ref{P2phi}) we can see that the large contributions to matrix
elements of the operators $Q_i^{(\prime)}$, i=11,...,16  arise
from penguin contractions with b quark in the loop. It is the
contributions which make the effects of Higgs mediated mechanism
sizable at the $\alpha_s$ order. At the $\mu=m_w$ scale only non
zero Wilson coefficients are $C_{11,13}^{(\prime)}$, and
$C_{15,16}^{(\prime)}(\mu)$ are obtained from
$C_{13}^{(\prime)}(m_w)$ due to the operator mixing under
renormalization. Moreover, because of the mixing of $Q_i$,
i=13,...,16, onto $Q_{7\gamma,8g}$, $C_{8g}(\mu)$ ($\mu\sim m_b$)
can be significantly enhanced\cite{hk}, which can have large
effects on both Br and $S_{\phi K}$, as can be seen from the SM
decay amplitude of $B\rightarrow \phi K_s$. %The mixing leads to
%\bea C_{8g}(\mu)&=&A(\mu_0)(\eta(\mu)^{\gamma_{cc}/2\beta_0}-
%\eta(\mu)^{\gamma_{8g8g}/2\beta_0})+
%C_{8g}(\mu_0)\eta^{\gamma_{8g8g}
%/2\beta_0},\\
%A(\mu_0)&=&\gamma_{a8g} V^{-1}_{ac}V_{cb}C_b(\mu_0)/(\gamma_{cc}
%-\gamma_{8g8g})\\
%\eta&=&\alpha_s(\mu_0)/\alpha_s(\mu),\\ a,b,c=1,2,3,4 \eea Here
%$\gamma_{aa}= diagonazed\,\, eigenvalues$ and V is the
%transformation matrix.

\begin{figure}\label{fig1}
{\includegraphics[width=4cm] {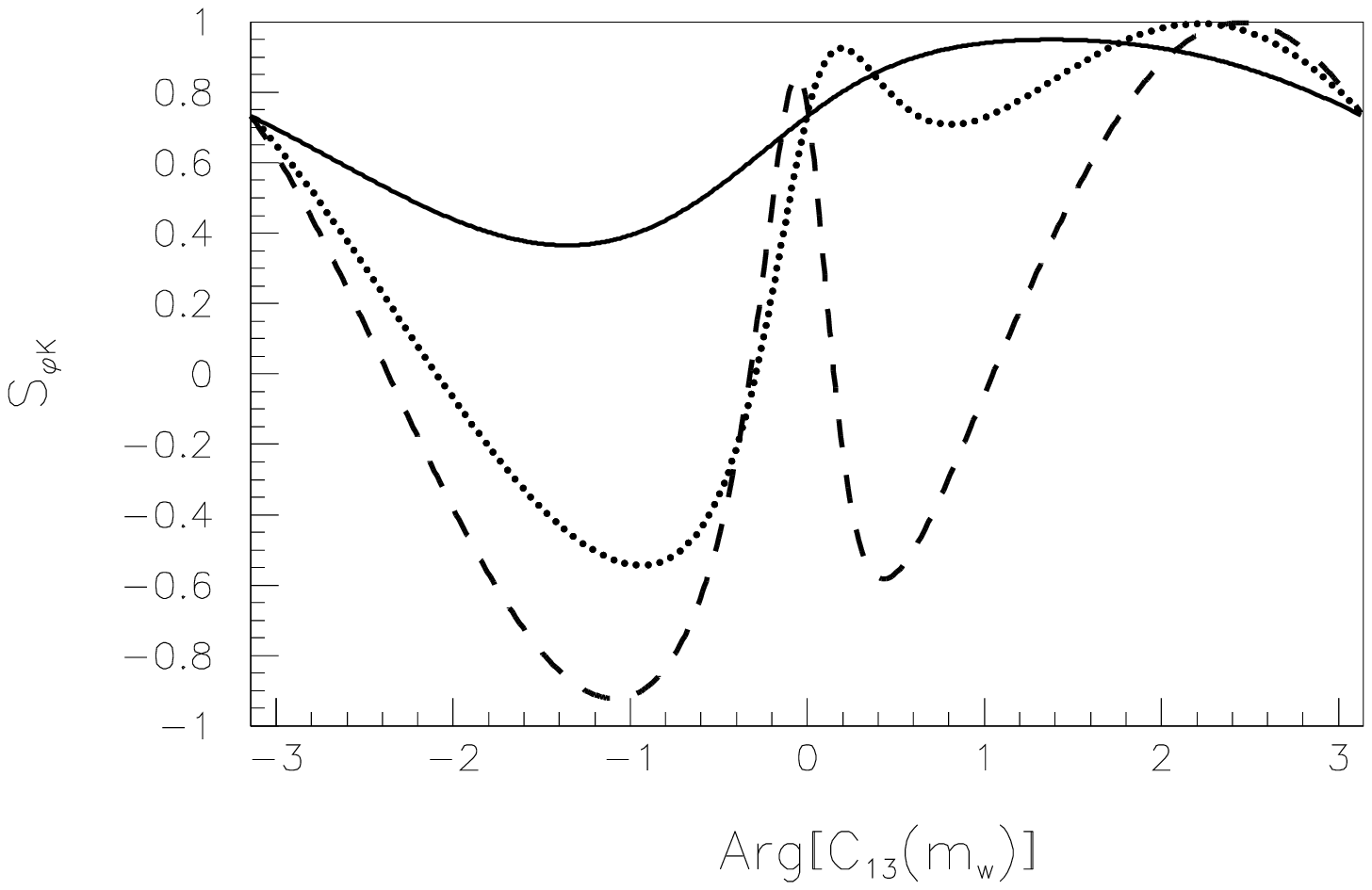}}
{\includegraphics[width=4cm] {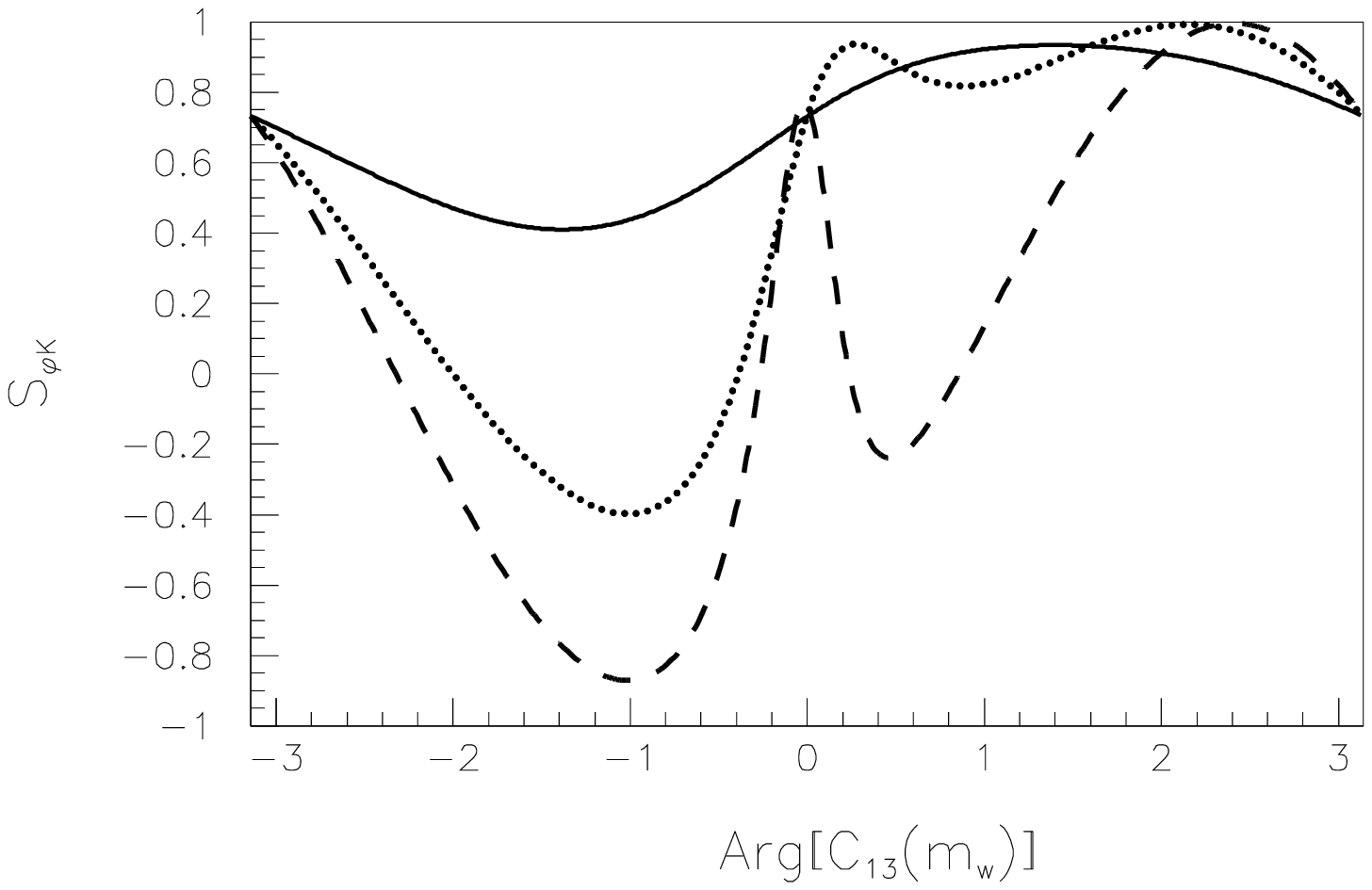}}
 \caption{
$S_{\phi K}$  versus the phase of $C_{13}(m_w)$ for fixed
$|C_{13}(m_w)|$ for (a) $\mu=m_b$, (b) $\mu= m_b/2$. The solid,
dotted and dashed curves correspond to $|C_{13}(m_w)|$= 0.05, 0.15
and 0.25 respectively.
 }
\end{figure}
In numerical calculations we use the SM decay amplitude of
$B\rightarrow \phi K_s$  given in Ref.\cite{bbns2}. We calculate
the evaluation of Wilson coefficients to the next-to-leading for
the SM operators and to the leading order for $Q_{11,...,16}$ and
their mixing with the SM operators. In order to see how large the
effects of NHBs can be we switch off all other SUSY contributions
except for those coming from NHB mediated
penguin operators. %\footnote{Indeed, as shown in Ref.~\cite{kane},
%qualitatively there are substantial cancellations between
%$C_{3,...,6}^{SUSY}$ and $C_{8g}^{SUSY}$ and the LL and RR
%insertions are significantly diluted.}.
We show model-independently in FIG. 1 $S_{\phi K}$ versus the
phase of $C_{13}(m_w)$ with fixed values of $|C_{13}(m_w)|$,
setting $C_{11}(m_w)$ and $C_{11,13}^\prime(m_w)$=0 in order to
see the essential feature of the $\alpha_s$ order corrections due
to the Higgs mediated mechanism. We find that $S_{\phi K}$ can
reach a minus value if $|C_{13}(m_w)|$ is equal to or larger than
$0.10$ for $\mu=m_b$ and $0.11$ for $\mu=m_b/2$. $S_{\phi K}$ can
be negative in quite a large range of the phase of $C_{13}$ for a
larger value of $|C_{13}(m_w)|$. $S_{\phi K}$ is dependent of but
not sensitive to the characteristic scale $\mu$ of the process.
For a specific model, e.g., MSSM which in the letter we are
interested in, we calculate all relevant Wilson coefficients
$C_{11,13}^{(\prime)}(m_w)$ in reasonable regions of parameter
space with middle and large $\tan\beta$ and present numerical
results in FIG. 2 below.

\begin{figure}\label{fig2}
{\includegraphics[width=4cm] {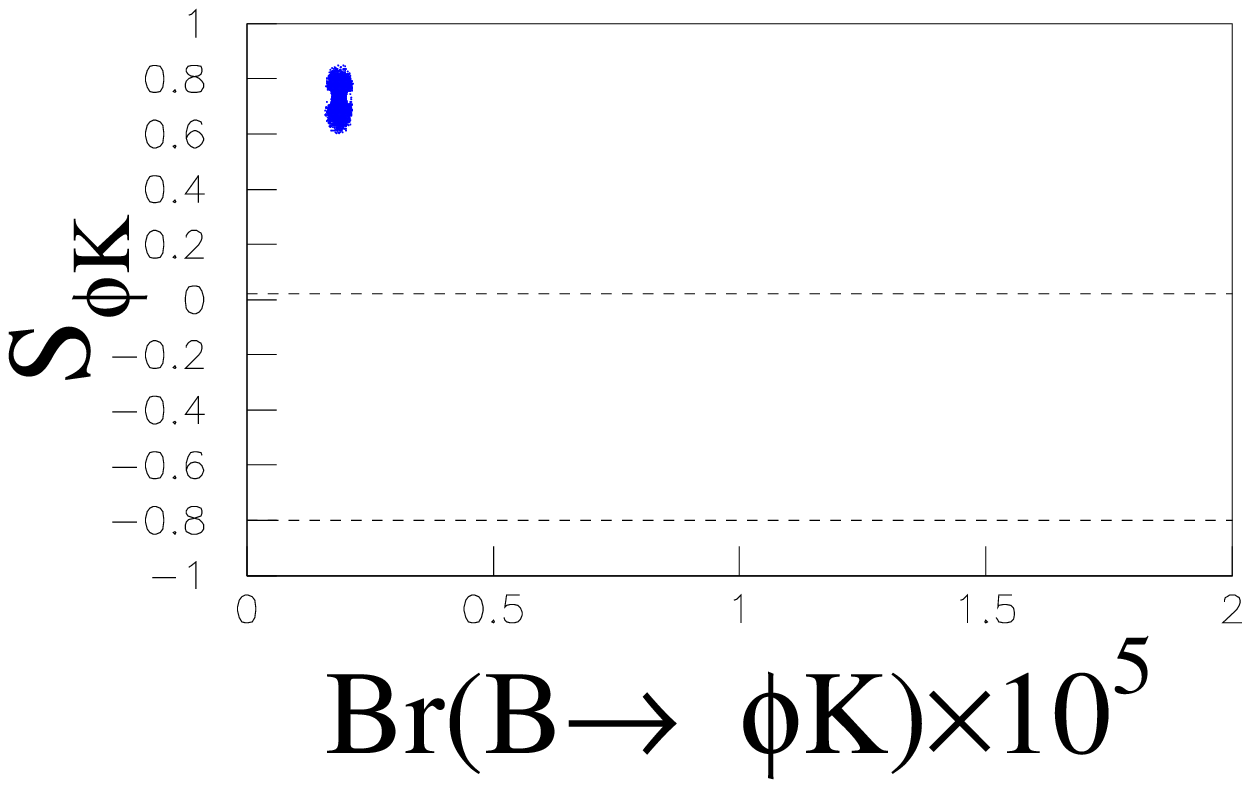}}
{\includegraphics[width=4cm] {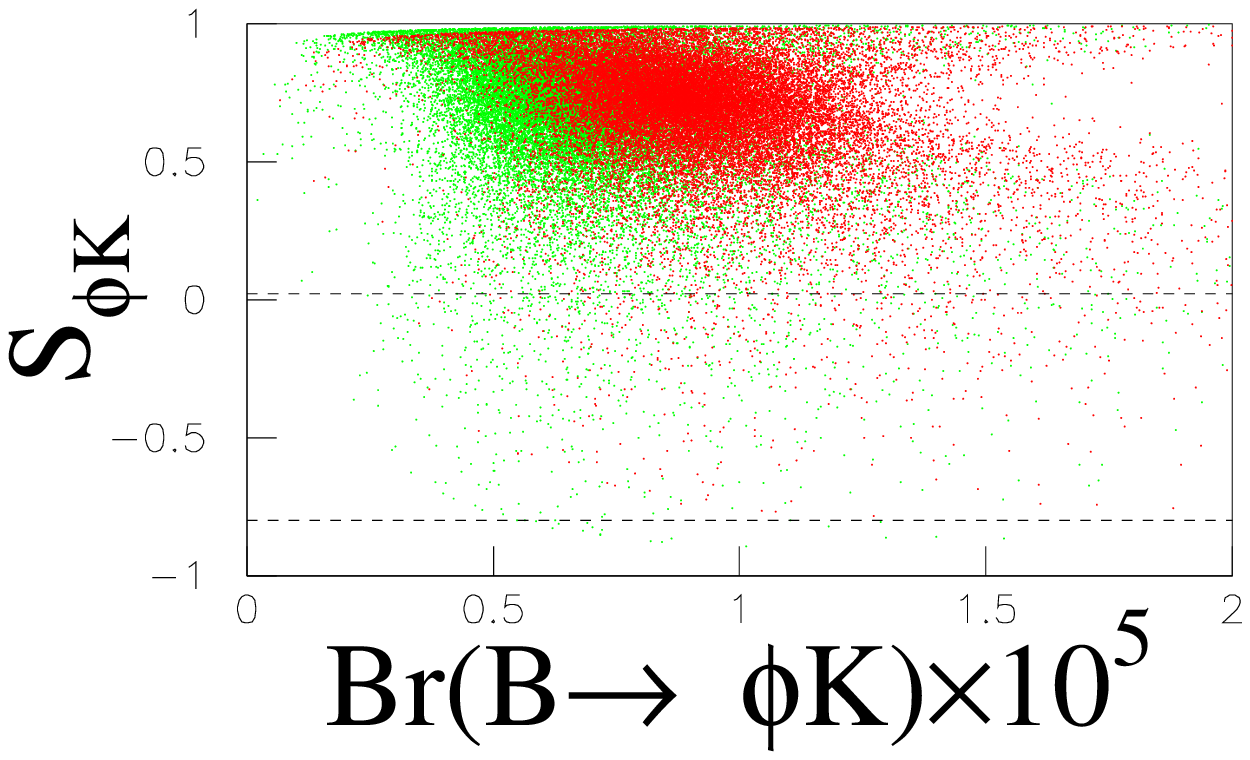}} \caption{
  The correlation between $S_{\phi K}$ and Br($B\to
\phi K_s$) for including only SM and NHB contributions. (a) is the
tree level result for $\mu=m_b$. (b) is including the $\alpha_s$
corrections. In Fig. 2b green and red dots denote the results for
$\mu=m_b$ and $\mu=m_b/2$ respectively. Current $1\sigma$ bounds
are shown by the dashed lines. }
\end{figure}

We impose two important constraints from $B\to X_s \gamma$ and
$B_s\to \mu^+\mu^-$. Considering the theoretical uncertainties, we
take $2.0\times 10^{-4} < {\rm Br}(B\to X_s \gamma)< 4.5\times
10^{-4}$, as generally analyzed in literatures. The current
experimental upper bound of ${\rm Br}(B_s\to \mu^+\mu^-)$ is
$2.6\times 10^{-6}$~\cite{bsmu}. Because the bound constrains
$|C_{Q_i}-C_{Q_i}^\prime|$ (i=1,
2)\footnote{$C_{Q_{1,2}}^{(\prime)}$ are the Wilson coefficients
of the operators $Q_{1,2}^{(\prime)}$ which are Higgs penguin
induced in leptonic and semileptonic B decays and their definition
can be found in Ref.~\cite{hy}. By substituting the quark-Higgs
vertex for the lepton-Higgs vertex it is straightforward to obtain
Wilson coefficients relevant to hadronic B decays.} we can have
values of $|C_{Q_i}|$ and $|C_{Q_i}^\prime|$ larger than those in
constrained MSSM (CMSSM) with universal boundary conditions at the
high scale and scenarios of the extended minimal flavor violation
in MSSM in which $|C_{Q_i}^\prime|$ is much smaller than
$|C_{Q_i}|$. We require that predicted Br of $B\to X_s \mu^+\mu^-$
falls within 1 $\sigma$ experimental bounds. We also impose the
current experimental lower bound $\Delta M_s > 14.4 ps^{-1}$ and
experimental upper bound ${\rm Br} (B\to X_s g)< 9\%$. In
numerical analysis we fix $m_{\tilde g}= m_{\tilde q}= 400 {\rm
GeV}$ and $\tan\beta=30$. We vary the NHB masses in the ranges of
$91 {\rm GeV} \leq m_h \leq 135 {\rm GeV}, 91 {\rm GeV} \leq m_H
\leq 200 {\rm GeV}$ with $m_h < m_H$ and $200 {\rm GeV} \leq m_A
\leq 250 {\rm GeV}$ for the fixed mixing angle $\alpha=0.6, \pi/2$
of the CP even NHBs and scan $\delta^{dAA}_{23}$ in the range
$|\delta^{dAA}_{23}| \leq 0.1$ which arises from the constraints
of $b\to s \gamma$ and $\Delta M_s$ (A = L, R). Here
$\delta^{dAA}_{23}$ is the parameter in the usual mass insertion
approximation~\cite{mia} and its definition can be found in
Refs.~\cite{hw,mia}. We perform calculations of hadronic matrix
elements of the Higgs penguin operators $Q_i^{(\prime)}$ at the
$\alpha_s^0$ order (tree level), i.e, in the naive factorization
formalism,  and to the $\alpha_s$ order. The results for tree
level (i.e., the naive factorization) and to the $\alpha_s$ order
are shown in FIGs. 2a and 2b, respectively. Fig. 2a is given for
$\mu=m_b$. In Fig. 2b green and red dots denote the results for
$\mu=m_b$ and $\mu=m_b/2$ respectively. We find that although
$S_{\phi K}$ can have a sizable deviation from SM but it can not
be smaller than 0.6 if one use the naive factorization to
calculate hadronic matrix elements. However, $S_{\phi K}$ for both
$\mu=m_b$ and $\mu=m_b/2$ can have a significant deviation from SM
when we include the $\alpha_s$ corrections. %The deviations are larger for
%tree level results than results including the $\alpha_s$
%corrections because of partial cancellations between the new
%contributions to the decay amplitude arising at tree level and
%those arising from the $\alpha_s$ corrections.
We find that there are regions of parameters where $S_{\phi K}$
falls in $1 \sigma$ experimental bounds and Br is smaller than
$1.6\times 10^{-5}$. %For $\mu= 2 m_b$ negative $S_{\phi K}$ can
%also be obtained.
The FIGs. 2a and 2b are plotted for $\alpha=\pi/2$. The similar
figures are obtained for $\alpha=0.6$. We stress that at tree
level $S_{\phi K}$ can have a sizable deviation from SM because we
can have values of $|C_{Q_i}|$ and $|C_{Q_i}^\prime|$ larger than
those in CMSSM with universal boundary conditions at the high
scale and the extended MFV scenarios in MSSM by a factor of 3 or
so, as pointed out above.  It is instructive to note that because
the LR and RL mass insertions do not contribute to the Higgs
penguin operators, the Higgs mediated mechanism probes the LL and
RR insertions, in contrast with the QCD and chromomagnetic
penguins whose effects are too small to alter $S_{\phi K}$
significantly in the case of only LL and RR
insertions~\cite{kane}\footnote{It is possible that effects of the
QCD and chromomagnetic penguins are significant to alert $S_{\phi
K}$ greatly for only a LL or RR insertion because there can be an
induced LR or RL insertion if $\mu \tan\beta$ is large
enough~\cite{kane}.}. We have also carried through the
calculations in the middle $\tan\beta$ case, $\tan\beta=8$, within
the range $|\delta^{dAA}_{23}|\leq 1$ and the results are similar
to those in the large $\tan\beta$ case.

Next we consider the case of large $m_A$. In the limit $M^2_{A}
\gg M^2_Z$ (for $M_{A} \geq 300$GeV), the charged, heavy CP-even
and CP-odd neutral Higgs bosons are nearly mass degenerate,
$M_{H^\pm} \simeq M_{H} \simeq M_{A}$, $\sin(\beta-\alpha)$
approaches $1$, and the properties of the lightest CP-even neutral
Higgs boson $h$ are almost identical to those of the SM Higgs
boson (so called the decoupling limit). In this case,
$C_{13}^{(\prime)}$ approaches to zero if $m_b A_b\ll m_{\tilde
g}^2$. So $S_{\phi K}$ for $\mu\sim m_b$ can not be smaller than
0.7 under the constraint from $B_s\to \mu^+\mu^-$. However, as
shown in Ref.~\cite{hwhiggs}, the decoupling limit can be relaxed
in some regions with $M^2_{A} \gg M^2_Z$ of the parameter space
which are allowed by experiments of $(g-2)_{\mu}$, $b\rightarrow s
\gamma$ and lower bounds of Higgs and sparticle masses, due to the
large off-diagonal scalar top and scalar bottom mass matrix
elements contributing to the Higgs sector by radiative
corrections. In the regions, the charged, heavy CP-even and CP-odd
Higgs bosons are not mass degenerate, and $\sin^2(\beta-\alpha)$
can be damped from $1$. We take two typical set of parameters in
the regions given in the TABLE 1 Case B of Ref.\cite{hwhiggs} to
calculate Wilson coefficients $C_i^{(\prime)}(m_w)$, $i=11, 13,
7\gamma $. Corresponding numerical results  are  that $S_{\phi K}
=-0.53$ and $-0.19$, ${\rm Br} = [0.77$ and $0.98] \times
10^{-5}$,
%-0.33$ and $-0.19, -0.05$, ${\rm Br} = [0.77, 1.04]\times 10^{-5}$
%and $[0.98, 1.29] \times 10^{-5}$ for $\mu=m_b$ and $m_b/2$
respectively.
%with complex $\delta^{dLL}_{23} = (0.133,0.333), (-0.0248,0.377)$,
%$\delta^{dRR}_{23} = (0.157,0.325), (-0.0156, 0.338)$.

\begin{figure}\label{fig3}
{\includegraphics[width=4cm] {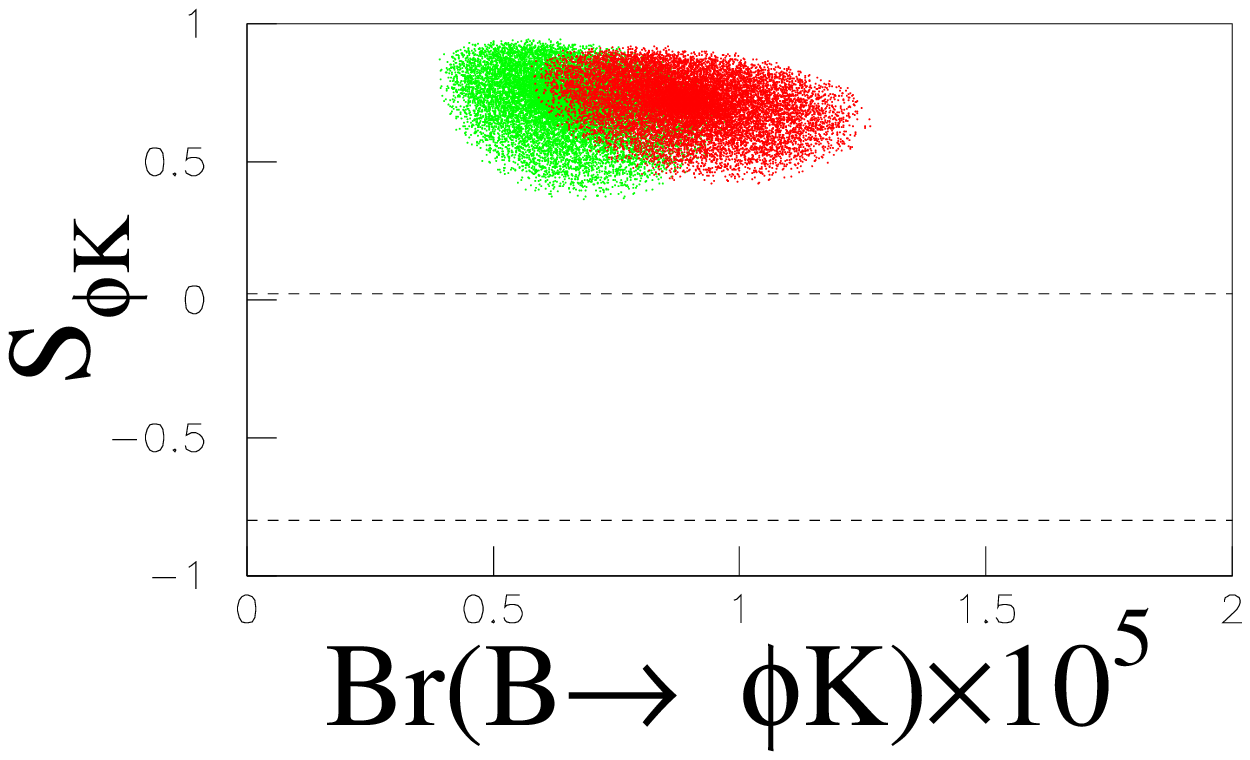}}
{\includegraphics[width=4cm] {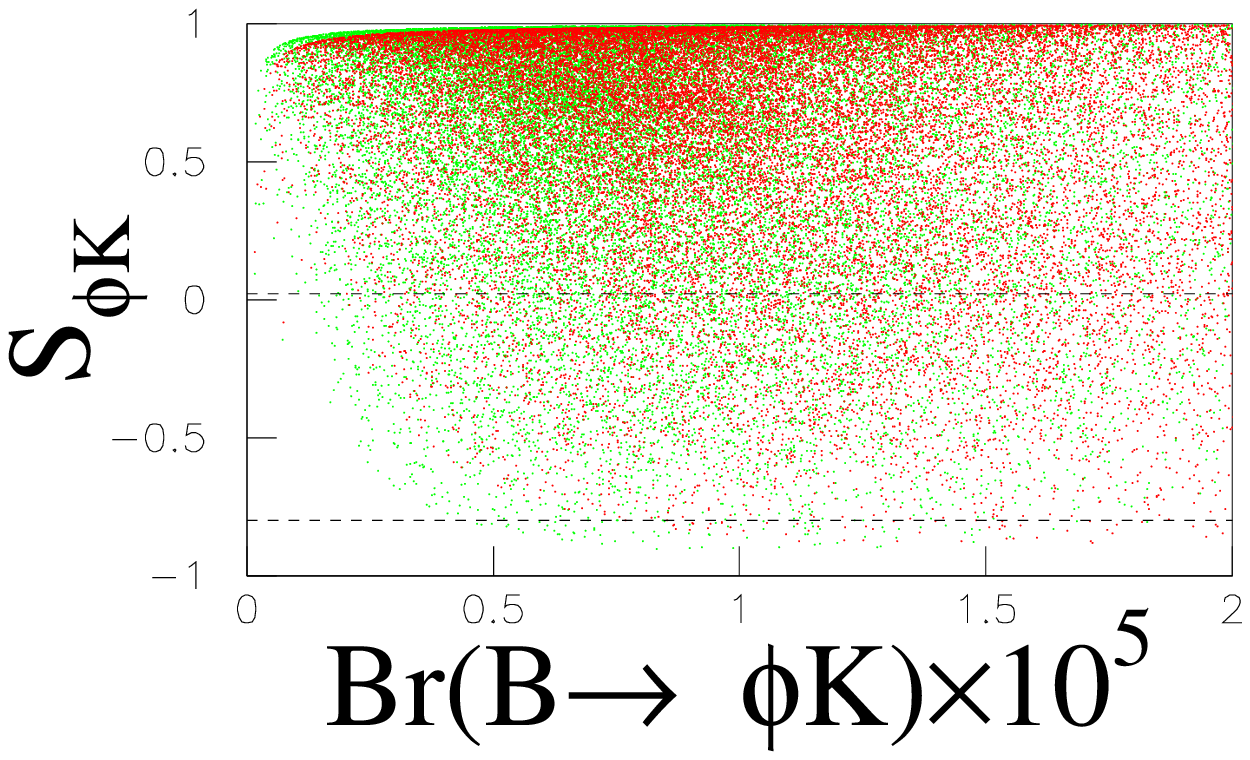}}
 \caption{
  The correlation between $S_{\phi K}$ and Br($B\to
\phi K_s$)  for (a) including only SM and NHB contributions and
only a LL insertion, (b) all contributions in MSSM. Green and red
dots denote the results for $\mu=m_b$ and $\mu=m_b/2$
respectively. Current $1\sigma$ bounds are shown by the dashed
lines. }
\end{figure}
We now consider the case of only a LL insertion. Then
$C_{11,...,16}^\prime=0$ and consequently the experimental bound
of $B_s\to \mu^+\mu^-$ constrains $|C_{Q_i}|$ (i=1, 2). We scan
$\delta^{dLL}_{23}$ in the range $|\delta^{dLL}_{23}| \leq 0.1$
with other parameters same as above under all relevant
constraints. The result is shown in Fig. 3a. We find that $S_{\phi
K}$ can significantly deviate the SM value but it can not be
smaller than 0.4. As pointed out above, if using the naive
factorization of hadronic matrix elements, the minimal value of
$S_{\phi K}$ is o.7, in contrast with that of including the
$\alpha_s$ corrections of hadronic matrix elements. The similar
result is obtained for the case of only a RR insertion.

Finally we switch on other SUSY contributions of which the
contributions of the chromomagnetic penguin with gluino-down type
squark in the loop are dominant. We scan $|\delta^{dLL,RR}_{23}|
\le 0.1$, with values of other parameters the same as those above
under the constraints from $b\to s \gamma$ and $\Delta M_s$ as
well as other relevant experiments mentioned above. The result
including both the chromomagnetic and QCD penguin contributions
and NHB penguin contributions is shown in FIG. 3b. One can see
from the figure that there are regions of parameters where
$S_{\phi K}$ falls in $1 \sigma$ experimental bounds and Br is
smaller than $1.6\times 10^{-5}$. Comparing FIG. 3b with FIG. 2b,
one can see that the regions of parameter space in the case
including both contributions are larger than those when including
only the NHB contributions.

Our results show that both the Br and $S_{\phi K}$ of $B\to \phi
K_S$ are dependent of the scale $\mu$ , which can be understood as
follows. Let us look at, e.g., the contributions coming from
$Q_{13,...,16}$ in $a_4^{neu}$ (see, Eq. (\ref{P2phi})), we have:
\begin{eqnarray}
&&{da_4^{neu}\over d\ln\mu}\nonumber\\
 &&= {d \left[ {\alpha_s\over 4 \pi }{C_F\over 4N_c}
(C_{13}\, 2\ln \mu  -4 C_{15}\, 2\ln \mu  -8 C_{16}\, 2\ln \mu
)G^0_\phi \right]
  \over d\ln \mu }+ ... \nonumber\\
 &&= {\alpha_s\over 4 \pi }{C_F\over 4N_c}
\big[(C_{13}\, 2  -4 C_{15}\, 2  -8 C_{16}\,2) )\big]G^0_\phi +
\alpha_s^2 \ln\mu\ {\rm terms}+ ... ,\label{mudep}
 \eea
 where "..." denotes the terms coming from $Q_{11}$ and $Q_{i}^
 \prime$, i=11,13,...,16.
 The first term  in Eq. (\ref{mudep}) is cancelled by
 \bea
 &&{d(-2 C_{8g}G^0_\phi)\over d\ln\mu}\nonumber\\&&={\alpha_s\over
 4 \pi }{C_F\over 4N_c}\big[
-2(C_{13}  -4 C_{15}  -8 C_{16})\big]G^0_\phi+ ...,\nonumber\eea
where $-2 C_{8g} G^0_\phi$ is the contribution coming from the
chromomagnetic dipole operator and "..." denotes the terms
independent of $C_{13,...,16}$. However, the $\alpha_s^2 \ln\mu\
{\rm terms}$ in Eq. (\ref{mudep}) is left. Although their
appearance is of $\alpha_s^2$ order, their magnitude indeed are of
$\alpha_s$ order. Similar analysis can be down for the other terms
in $a_4^{neu}$.

It is necessary to make a theoretical prediction in SM as
precision as we can in order to give a firm ground for finding new
physics. For the purpose, we calculate the twist-3 and weak
annihilation contributions in SM using the method in
Ref.~\cite{ch} by which there is no any phenomenological parameter
introduced\footnote{The twist three and annihilation contributions
are included by Cheng and Yang in Ref.\cite{hmw}. However, they
introduce a phenomenological parameter instead of the integral
containing end point singularity and consequently the prediction
ability of theory is decreased.}.  The numerical results show that
the annihilation contributions to $Br(B\to \phi K)$ are of order
$10^{-8} $ which are negligible, the twist-3 contributions to Br
are also very small, smaller than one percent, and both the
annihilation and twist-3 contributions to $S_{\phi K}$ are
negligible.

% \footnote{Because our numerical results for
%$f_{II}=4.5+1.4 I$ is smaller than the corresponding value
%estimated by Cheng {\it et\ al}\cite{cy} and the non-factorizable
%diagrams give opposite contributions to the branching ratio of
%$B\to \phi K$, the numerical result for
% $Br(B \to \phi K)$ is about 15 percent larger than
%theirs' $3.6\times 10^{-6}$, with the other condition same.}.

%We did not calculate the direct CP asymmetry of $B\to \phi K_S$
%although it is straightforward to calculate. The reason is that at
%present there are large uncertainties in theoretical calculations
%of strong phases in SM and there are a significant disagreement of
%calculated results in literatures~\cite{bbns2,lhn}. It is not
%convincing to say something on new physics from the direct CP
%asymmetry in the decay based on present calculations of strong
%phases in SM.

In summary, we have shown that the NHB contributions in MSSM can
have a significant effect on both the Br and $S_{\phi K}$ of $B\to
\phi K_S$. We can have large Wilson coefficients (comparable with
those in SM) of the NHB penguin induced operators
$Q_i^{(\prime)}$, in some regions of the parameter space in MSSM
under all relevant experimental constraints, which leads to that
even at the tree level of hadronic matrix element calculations the
time dependent CP asymmetry $S_{\phi K}$ can deviate from SM
sizably. Due to the large contributions to the hadronic elements
of the operators at the $\alpha_s$ order arising from penguin
contractions with b quark in the loop, both the Br and $S_{\phi
K}$ are sizably different from those in SM. The mixing
$Q_{13,...,16}$ with $Q_{8g}$ makes the Wilson coefficient
$C_{8g}(\mu)$ ($\mu\sim m_b$) enhanced a lot, which has large
effects on both $S_{\phi K}$ and the Br. In some regions of the
parameter space in MSSM, the NHB contributions alone can lead to
negative $S_{\phi K}$ which is reported by BaBar and Belle in 2002
and even make $S_{\phi K}$ agreed with 2003
Belle data in a $1 \sigma$ experimental bounds. %We
%have also studied the realistic case in which the QCD and
%chromomagnetic penguins and NHB penguins co-exist. In this case
%$S_{\phi K}$ can be negative in regions of parameter space larger
%than those in the case of only one mechanism making play.
Our results show that both the Br and $S_{\phi K}$ of $B\to \phi
K_S$ are dependent of the scale $\mu$. The sizable scale
dependence, which implies hadronic uncertainties up to the
$\alpha_s$ order, comes mainly from the $O(\alpha_s)$ corrections
of hadronic matrix elements and also from leading order Wilson
coefficients $C_i^{(\prime)}$, i=11,...,16. However, despite there
are hadronic uncertainties, the conclusion that the NHB
contributions in MSSM can have a significant effect on both the Br
and $S_{\phi K}$
of $B\to \phi K_S$ can still be drawn definitely. %Finally, we
%would like to point out that it is straightforward and interesting
%to study effects of the NHB mediated mechanism on decays $B\to
%\eta^{(\prime)} K$.

\section*{Acknowledgment}
The work was supported in part by
 the National Nature Science Foundation of China.

\end{document}